\documentclass[twocolumn,twoside,fleqn,showpacs,showkeys,pra,aps,10pt,superscriptaddress,final]{article}

\usepackage{hep}
\graphicspath{{figs/}}      

\def\refname{References}

\def\journalname{??}

\def\@pacs@name{PACS numbers: }%
\def\@keys@name{Keywords: }%

\def\Dated@name{Dated: }%
\def\Received@name{Received }%
\def\Revised@name{Revised }%
\def\Accepted@name{Accepted }%
\def\Published@name{Published }%
\def\address{\replace@command\address\affiliation}%
\def\altaddress{\replace@command\altaddress\altaffiliation}%

\definecolor{orangec}{cmyk}{.24,.91,.96,.18}
\definecolor{orangecc}{cmyk}{.24,.94,.96,.18}
\definecolor{oorangec}{cmyk}{.8,.2,.5,.4}
\definecolor{ooorangec}{cmyk}{1,.9,0.08,.04}      

\definecolor{orangec}{cmyk}{.15,.7,.96,.0}
\definecolor{orangecc}{cmyk}{.15,.7,.96,.0}

\newcommand{\ODDBOX}{\noindent\raisebox{-13.3cm}{\includegraphics{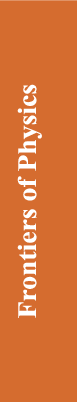}}}
\newcommand{\EVENBOX}{\noindent\raisebox{-13.3cm}{\includegraphics{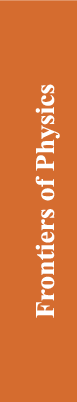}}}

\newfont{\yihao}{cmb10 at 18pt}
\newcommand{\Yihao}{\fontsize{18pt}{13.5pt}\selectfont}
\newcommand{\xiaosi}{\fontsize{11.5pt}{13.5pt}\selectfont}         

\newfont{\xbt}{cmb10 at 12pt}

\def\frontmatter@title@format{%
    \centering%
    \usefont{T1}{fradmcn}{m}{n}\yihao}%
\def\@keys@name{{\color{ooorangec}\bf Keywords~~}}%
\def\@pacs@name{{\color{ooorangec}\bf PACS numbers~~}\vspace{2mm}}%
\def\frontmatter@authorformat{\vspace{5mm}\centering\bf}%

\newcommand{\catchline}[2]{
	{\vspace*{-16.4mm}\small%
	\noindent #1\\%
	\noindent #2\\[-2mm]%
    {\color{orangec}{\rule{\textwidth}{.5pt}}}\\[2mm]
    {\color{orangec}{\Yihao\textbf{\textsc{\Papertype}}}}\\[6mm]
    }\relax\par
}

\newcommand{\doi}[1]{DOI {#1}}
\renewcommand{\title}[1]
{\vspace*{-5mm}\begin{center}
{\Yihao\bf #1}
\end{center}
}

\renewcommand{\author}[1]
{\vspace*{0mm}
\begin{center}
{\bf #1}
\end{center}
}
\newcommand{\add}[1]{\begin{center}{\small\it #1}\end{center}}

\newcommand{\abs}[1]{
\begin{center}
\parbox[t]{156mm}{\noindent\color{oorangec}#1}
\end{center}}

\newcommand{\keywords}[1]{
\begin{center}
\parbox[t]{156mm}{\noindent{\bf\color{ooorangec}Keywords}\ \ #1}
\end{center}}

\newcommand{\pacsnumbers}[1]{
\begin{center}
\parbox[t]{156mm}{\noindent{\bf\color{ooorangec}PACS numbers}\ \ #1\vspace*{5mm}}
\end{center}}

\newcommand{\acknowledgements}[1]{\vspace*{4mm}\noindent{\renewcommand{\baselinestretch}{1.05}\footnotesize{\color{ooorangec}\bf Acknowledgements}\quad{#1}}}

\def\journalname{??}
\def\volumenumber#1{\gdef\@volumenumber{#1}}%
\def\@volumenumber{}%
\def\issuenumber#1{\gdef\@issuenumber{#1}}%
\def\@issuenumber{}%
\def\volumeyear#1{\gdef\@volumeyear{#1}}%
\def\@volumeyear{}%

\renewcommand\thesection{\arabic{section}}
\renewcommand\thesubsection{\arabic{section}.\arabic{subsection}}
\renewcommand\thesubsubsection{\arabic{section}.\arabic{subsection}.\arabic{subsubsection}}

\titleformat{\section}[hang]{\color{ooorangec}\vspace*{-1.2mm}\titlerule\vspace{1mm}\large\usefont{T1}{fradmcn}{m}{n}\xbt}{\thesection}{1em}{}
\titlespacing{\section}{0mm}{8mm}{5mm}

\titleformat{\subsection}{\normalfont\normalsize\color{ooorangec}}{\thesubsection}{1em}{}
\titlespacing{\subsection}{0mm}{5mm}{3mm}

\titleformat{\subsubsection}{\normalfont\normalsize\it\color{ooorangec}}{\thesubsubsection}{1em}{}
\titlespacing{\subsubsection}{0mm}{3mm}{3mm}

\titlecontents{section}
[0em] {\normalfont} {\thecontentslabel\quad} {\thecontentslabel}
{\titlerule*[.7pc]{}\contentspage}

\titlecontents{subsection}
[1.5em] {\normalfont} {\thecontentslabel\quad} {\thecontentslabel}
{\titlerule*[.7pc]{}\contentspage}

\titlecontents{subsubsection}
[4em] {\normalfont} {\thecontentslabel\quad} {\thecontentslabel}
{\titlerule*[.7pc]{}\contentspage}

\usepackage[small]{caption2}

\newlength{\halfpagewidth}
\divide\halfpagewidth by 1

\usepackage{hyperref}
\usepackage{mathrsfs}
\usepackage{ulem}

\begin{document}

\newcommand{\Papertype}{\sc Research article} 
\def\volumeyear{2020} 
\def\volumenumber{15(5)} 
\def\issuenumber{54501} 
\def\journalname{Front. Phys.} 
\newcommand{\doiurl}{10.1007/s11467-020-0966-4} 
\newcommand{\allauthors}{Xiangru Li and Woliang Yu, Xilong Fan and G. Jogesh Babu} 

\twocolumn[
\begin{@twocolumnfalse}
\catchline{\journalname~\volumenumber,~\issuenumber~(\volumeyear)}{\doi{\doiurl}} 
\thispagestyle{firstpage}


\title{Some Optimizations on Detecting Gravitational Wave Using Convolutional Neural Network}

\author{Xiangru Li$^{1*}$, Woliang Yu$^{2}$, Xilong Fan$^{3**}$,G. Jogesh Babu$^{4}$}

\add{1. School of Computer Science, South China Normal University, Guangzhou 510631, China\\
2. School of Mathematical Sciences, South China Normal University, Guangzhou 510631, China\\
3. School of Physics and Technology, Wuhan University, Wuhan, Hubei 430072, China\\
4. Pennsylvania State University, University Park PA, 16802, USA\\
Corresponding author.\ E-mail: $^*$lixiangru@scnu.edu.cn, $^{**}$xilong.fan@whu.edu.cn\\
Received March 25, 2020; Accepted May 06, 2020}

\abs{This work investigates the problem of detecting gravitational wave (GW) events based on simulated damped sinusoid signals contaminated with white Gaussian noise. It is treated as a classification problem with one class for the interesting events. The proposed scheme consists of the following two successive steps: decomposing the data using a wavelet packet, representing the GW signal and noise using the derived decomposition coefficients; and determining the existence of any GW event using a convolutional neural network (CNN) with a logistic regression output layer. The characteristics of this work is its comprehensive investigations on CNN structure, detection window width, data resolution, wavelet packet decomposition and detection window overlap scheme. Extensive simulation experiments show excellent performances for reliable detection of signals with a range of GW model parameters and signal-to-noise ratios. While we use a simple waveform model in this study, we expect the method to be particularly valuable when the potential GW shapes are too complex to be characterized with a template bank.}


\pacsnumbers{04.30.Db, 07.05.Mh}

\vspace*{-6mm}

\end{@twocolumnfalse}
]

\section{
Introduction} \label{sec:intro}

Gravitational waves (GWs) potentially give a remarkable opportunity to penetrate unprecedented regions of the universe and compact astronomical objects. Direct detections of GW events by the advanced LIGO and Virgo detectors   \citep{Journal:Abbott:2016PRL061102,Journal:Abbott:2016PRX041015,Journal:Abbott:2017PRL221101,Journal:Abbott:2017PRL141101,Journal:Abbott:2017PRL161101}
   enhance our understanding of gravitational theory and bring gravitational-wave astronomy into an observational science \citep{Journal:AdrianMart:2016,Journal:Abbott:2017L13,Journal:Abbott:2017L12,Journal:Abbott:2017Nature}.
	
GW burst software pipelines, used by the LVC (LIGO-Virgo Collaboration) search for generic gravitational-wave transients, are implemented with minimal assumptions about the signal waveform, polarization, source direction, and occurrence time based on their time-frequency morphology \citep{Journal:Abbott:2017:PRD}. Although the main targets of the burst pipelines are unmodeled transient GW events, CBC (compact binary coalescence) signals are well-discovered by the event detection procedures \citep{Journal:Abbott:2016PRD122004}. The parameters of some characteristics of these unmodeled GW signals, such as the central frequency, duration and bandwidth, can be estimated from the time-frequency transform of the observed time series and then converted to physical parameters by some comparison-based schemes with sophisticated waveforms computed from astrophysical models \citep{Journal:Abbott:2016PRD122004}.
	
This work focuses on a procedure to detect the existence of any simulated damped sinusoid GW signals from time series with white Gaussian noises. This type waveform could be a burst of GWs from the ringdown of a perturbed black hole \citep{Journal:Vishveshwara:Nature1970}, from the post-merger of  a binary neutron star (BNS) system, or from excitation of fundamental modes in neutron stars \citep{Journal:Benhar:PRD2004}. However, real detector data contains non-Gaussian noise artifacts. Therefore, this work is a preliminary investigation which assumes that such non-Gaussian disturbances have been removed from the data or from the detector \citep{Journal:Powell:2015,Journal:Jade:2017, Journal:Zevin:2017}. The performance representativeness of this work on real detector data need further studies in future.

In GW detection, a benchmark method is the matched filtering \citep{Journal:Allen:2012,Journal:Cannon:2012,Journal:Babak:PRD2013,Journal:Usman:2016,Journal:Abbott:2016PRD122004}. Its typical limitation is the expensive computational requirements. Therefore, some investigations are made based on a machine learning scheme, convolutional neural networks (CNNs), for GW detection, and obtained some exciting results comparable to the matched filtering \citep{PRB:George:2018,Journal:Gabbard:2018PRL,FoP:Luo:2020,FoP:Lin:2020}.

Based on the knowledges of machine learning and signal processing, the practical performance of a detection system depends on the configurations of such factors as a detection window (duration of a time series under being processed), a sampling rate/data resolution. In related literatures on GW detection, however, the configurations of these factors weren't  investigated correspondingly. For example, the width of a detection window is 1 second in \citep{PRD:George:2018,PRB:George:2018,Journal:Gabbard:2018PRL}; Two sampling rates are 8192 HZ \cite{PRD:George:2018,PRB:George:2018,Journal:Gabbard:2018PRL} and 2048HZ \cite{NIPS:Gebhard:2017,Journal:Gebhard:PRD2019}. Therefore, this work studies the influences from these factors on GW detection and their optimization, and it is shown that the performance of the CNN-based scheme can be improved evidently by optimizing these factors.

Another interesting issue in this problem is the necessity of time-frequency analysis and selection of frequency analysis methods in CNN-based GW detection scheme. Because the gravitational wave bursts are short transients of gravitational radiation and their time-frequency information is a typical characteristic for GW candidates, time-frequency analysis is a proposed procedure in some GW detection studies based on non-machine learning scheme \citep{CQG:Chatterji:2004,Journal:Sutton:2010NJP,Journal:Abbott:2017:PRD} and many studies for glitch classification \citep{Journal:Zevin:2017,Journal:Mukund:2017,ArXiv:Sara:2017,PRD:George:2018CUC,IS:Bahaadini:2018}, for example, wavelet decomposition and Q-transform \citep{Journal:Brown:1991,CQG:Chatterji:2004}, coherent Waveburst (cWB) \citep{Journal:Klimenko:2008CQG29,Journal:Klimenko:2016PRD19,Journal:Abbott:2017:PRD} and
omicron-LIB (oLIB) \citep{Journal:Abbott:2017:PRD,Journal:Lynch:2017PRL18}, BayesWave (BW) \citep{Journal:Cornish:2015CQG30,Journal:Littenberg:2015PRD31,Journal:Abbott:2017:PRD}, X-Pipeline \citep{Journal:Chatterji:2006PRD,Journal:Sutton:2010NJP}.
However, the existing investigation on CNN-based GW detections are conducted by inputting a time series into a CNN network without investigations on the effects from this kind frequency procedures. Therefore, the necessity of time-frequency analysis and the affects from time-frequency analysis methods are still two open problems to be studied in CNN-based GW detection scheme.

Based on the above two considerations, this work proposed a GW detection scheme based on CNN and a time-frequency analysis method wavelet packet decomposition, and gave an optimized configuration of a detection window (width of a detection window, overlap length between two successive detection windows) and data resolution. It is shown that the proposed optimization of this work improved the GW detection sensitivity evidently, and the accumulation effects from such factors could't be ignored in GW detection.

\section{
Data}\label{sec:data}
Let $\boldsymbol{s}\in \mathbf{R}^D$ represent an observed data/sample, where $\boldsymbol{s} = \boldsymbol{n}$ or $\boldsymbol{s} = \boldsymbol{n} + \boldsymbol{h}$, $\mathbf{R}^D$ is a $D$-dimensional Euclidean space, $D$ is a positive integer denoting the dimension of a sample, $\boldsymbol{n}$ are some noises and $\boldsymbol{h}$ is a GW signal;  Suppose $y \in \{1, 0\}$ is a label of $\boldsymbol{s}$: $y=1$ indicates that there exists a GW signal in $\boldsymbol{s}$, and $y=0$ indicates that there isn't any GW signal in $\boldsymbol{s}$. For convenience, the cases $\boldsymbol{s} = \boldsymbol{n}$ and $\boldsymbol{s} = \boldsymbol{n} + \boldsymbol{h}$ are referred to as pure noise and signal with additive noise respectively.  This work is to establish a  model
	\begin{equation}\label{Equ:estimation_model}
	y=f(\boldsymbol{s})
	\end{equation}
using statistical theories to detect the existence of any GW in our observed data. This is referred to as a classification problem in machine learning community.
		
In this kind statistical classification scheme, the knowledge of GW detection is expressed using a set, $ S_{tr} = \{(\boldsymbol{s}_i, y_i), i= 1,\cdots,N \}$, of observed/theoretical data and their labels, where $\boldsymbol{s}_i \in \mathbf{R}^D$ is a pure noise sample or a signal with additive noise, $y_i \in \{1, 0\}$ is its label indicating whether there exists GW in $\boldsymbol{s}_i \in \mathbf{R}^D$. That is to say, the empirical data set $S_{tr}$ is a knowledge carrier. The model $y=f(\boldsymbol{s})$ is estimated from this empirical data set using some computational schemes. For conveniences, this estimation is denoted by $\hat{f}$. The empirical set $ S_{tr}$ is referred to as a \textbf{training set}. The empirical data $(\boldsymbol{s}, y)$ is called a sample.
	
To evaluate the performance of $\hat{f}$, some samples independent of $S_{tr}$ are needed. These evaluation samples constitute a set $S_{te}$, which is referred to as a \textbf{test set}. The independence here is to ensure the objectivity of the model evaluation.

One typical characteristic of the problem is that the GW is usually very weak and the observed signal is contaminated with various disturbances. All of the existing disturbances are collectively referred to as noises. Actually, \citet{Journal:Gabbard:2018PRL}, \citet{PRD:George:2018}, and this work are of investigating the feasibility of detecting GW using a machine learning scheme and its potential. In these works, a fundamental assumption is  that glitches have been removed from the data or from the detector \citep{ArXiv:Sara:2017,Journal:Powell:2015,Journal:Jade:2017, Journal:Zevin:2017, Journal:Mukund:2017,Journal:Bose:PRD2016}. Therefore, this work modeled the noises approximately using a series of independent Gaussian random variables with zero mean , without any considerations of influences from glitches, other transient sources of detector noise. That is to say, the assumed noise model is a white Gaussian noise. In real applications, however, the intensity of noise is non-stationary and the signal-to-noise ratios (SNRs) on different segments of observed data are inconstant. To simulate these cases, this work computed Gaussian noises with their corresponding variance depending on randomly chosen SNRs (Table \ref{Tab:DataSets}). Therefore, the noises are non-stationary in this work. These are of earlier stage investigations, and we will study the performance of this kind scheme based on real detector noise in future works.

This work uses three data quality measures: the optimal SNR $\rho_{opt}$ \citep{Journal:Gabbard:2018PRL},
the ratio between the maximum amplitude of the waveform and the white noise standard deviation $A^{S/N}$,
and the matched filtering SNR $\texttt{SNR}_{mf}$ \citep{Journal:Owen:1999}
	\begin{equation}\label{Equ:SNR:rho}
		 \rho_{opt}=2\sqrt{\int_{f_{min}}^{\infty}\frac{|\tilde{h}(f)|^2}{\hat{S}_s(f)}df},
	 \end{equation}
	\begin{equation}\label{Equ:SNR}
		 A^{S/N}={\max({\boldsymbol{h}})}/{\sigma(\boldsymbol{n})},
	 \end{equation}
	\begin{equation}\label{Equ:SNR:SNRmf}
         \texttt{SNR}_{mf} = \frac{<s, h>}{\sqrt{<h,h>}}
	 \end{equation}
where
	\begin{equation}\label{Equ:SNR:SNRmf:innerProduct}
		 <s, h> = 4~Re{\int_{0}^{\infty}\frac{\tilde{s}^{*}(f) \tilde{h}(f)}{\hat{S}_{s}(f)}df},
	 \end{equation}
$\boldsymbol{h}(t)$ is a signal/template, $\tilde{\boldsymbol{h}}(f), \tilde{a}(f)$ and $\tilde{b}(f)$ are the Fourier transform of $\boldsymbol{h}(t)$, $a(t)$ and $b(t)$ respectively,  $\tilde{a}^{*}(f)$ is the complex conjugate of $\tilde{a}(f)$, $\hat{S}_{s}(f)$ is the power spectral density (PSD) estimated using the Welch's method \citep{ITAE1967:Welch,Journal:Meadors:2014,Matlab:Welch}, $f_{min}$ is the frequency at which we start to accumulate SNR, and $\sigma(\boldsymbol{n})$ is the standard deviation of noise $\boldsymbol{n}$. The $\texttt{SNR}_{mf}$ and $\rho_{opt}$ are commonly used measures in gravitational wave community and their usages are helpful in comparing experimental results with related literatures. The $A^{S/N}$ also makes the data simulation easy.

In this work, the GW time series is computed using a damped sinusoid waveform \citep{Abadie:2014}:
	\begin{equation}\label{Equ:GW_Model}
	\boldsymbol{h}(t|t_0) = \left\{\begin{array}{ll}
	0                                                &t<t_0\\
	e^{-\frac{t-t_{0}}{\tau}}\sin(2\pi f_0(t-t_{0})) &t\geq t_0,
	\end{array}
	\right .
	\end{equation}
\noindent where $t_{0}$ is the occurrence time of a GW, $f_0$ a central frequency and $\tau = Q/(\sqrt{2}\pi f_0)$ a decay parameter. The frequency $f_0$ covers a range [40, 200] Hz uniformly, and  quality factors $Q$ randomly  takes three values 3, 9 and 100 \citep{Abadie:2014} as the {\it fiducial}  models  in simulated data (this work also tested the detection sensitivity on more $Q$ values).

In a supervised learning scheme, a training set and a test set are two requisites. The training set is a knowledge carrier in this kind GW detection scheme and a proposed scheme should be established by learning from them. The test data set acts as a referee to be used in objectively evaluating the performance of the established model. In application, however, it is difficult to make sure that training data and test data accurately share the same noise contamination, occurrence time $t_0$, $f_0$ or SNR. Therefore, each training data and test data are computed based on randomly and independently generated noise, $t_0$, $f_0$ and SNR in this work.

To evaluate the performance comprehensively and illustrate the methods optimizing the proposed scheme, this work constructs a series of training sets and test sets with various parameter configurations. More about the purposes can be found in section \ref{sec:Experiments}. To be clearly, configurations of these data sets are presented in Table \ref{Tab:DataSets}.

In each practical case, a GW detection scheme only can learn from a limited number of training data/templates, whose parameters cover a limited number of values. For example, in this work, the training set$^1$ covers 25000 cases of $t_0$ and SNR. It is impossible to enumerate all possible configurations in a training set. In application, however, there are infinite number of potential configurations of parameters, e.g., occurrence time $t_0$, $f_0$, $A^{S/N}$, even in a restricted ranges. The potential cases to be dealt with should have more diversity than that of training data in both theory and application. Therefore, some data with different parameter configurations from that of training data should be processed satisfactorily. To evaluate this performance as good as possible, this work used much more test data than training data, and the parameter ranges of the test data are larger than that of training data.

\begin{table}\scriptsize
\centering
\caption{Configurations of data sets used in this work. Occurrence time $t_0$ of a GW signal is unknown beforehand, therefore $t_0$ is randomly generated from a given time window and is different from sample to sample.  The data sets are computed from the model (\ref{Equ:GW_Model}). s: second, NS: number of samples, NS\{m,n\}: m samples consisting of Gaussian noise and n samples consisting of GWs injected in Gaussian noise. In this table, a range [a, b] of $t_0$ refers to $t_0$ being a random number greater than or equal to a seconds and less than or equal to b seconds for each signal in corresponding data set; the ranges of $A^{S/N}$ is defined similarly.  }
\begin{tabular}{|c|c|c|c|c|}
  \hline
  Data Sets         & $A^{S/N}$   & $t_0$ (s)   &NS        &NS\{m,n\}\\ \hline
  Training set$^1$ & [0.2, 0.8]  &[0, 0.03]             & 50000    & \{25000,  25000\}   \\ \hline
  Test set$^1_1$   &  [0.15, 1.05]&[0, 0.026]            & 190000   & \{95000,  95000\}   \\ \hline
  Training set$^2$ & [0.2, 0.8]  &[0, 1]                & 50000    & \{25000,  25000\}   \\ \hline
  Test set$^2_1$   & [0.15, 1.05]&[0, 1]                & 190000   & \{95000,  95000\}   \\ \hline
\hline
\end{tabular}\label{Tab:DataSets}
\end{table}

\begin{figure}
		\centering
\includegraphics[height=5cm]{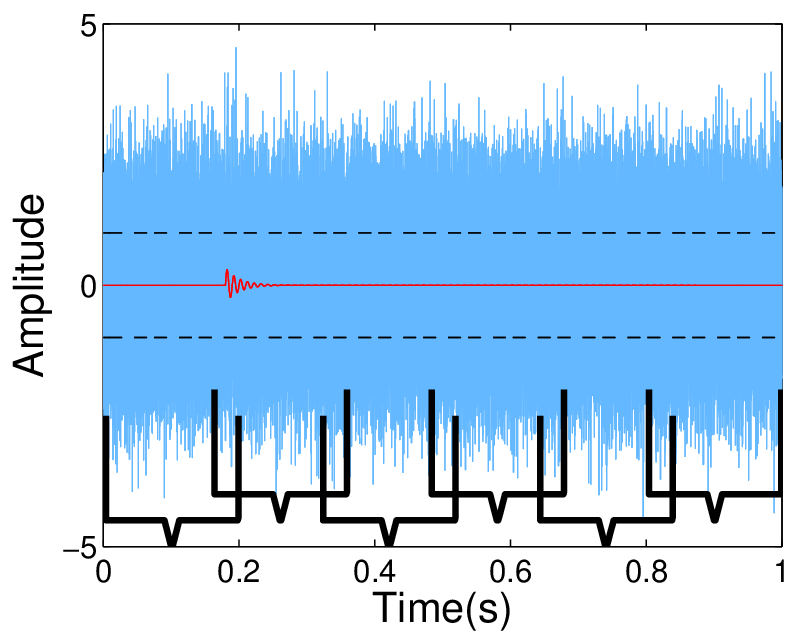}
		\caption{
A sample of simulated data. In this figure, the blue/outer curve is a sample to be analyzed, which consists of GW signal and noises. The red/inner curve is the theoretical GW signal. The $A^{S/N}$ of this work covers a range [0.15,1.05]. For visibility, this figure shows a sample with $A^{S/N}$$= 0.3$. In application, data is obtained in stream. The GW detection system needs to move the detection window on the stream to take the data segment from it to analyze. This diagram shows our proposed scheme for moving detection window (indicated with brackets). Two other parameters of this signal are $f_0$=100 Hz, Q=9.  Two horizonal and dashed lines indicate the standard deviation of noise.
		}
		\label{Fig:GW_flow}
	\end{figure}

	In applications, data are obtained continuously in a form of streams. A GW detection software pipeline needs to move a time window on the data stream, extract a data segment from every window, and determine whether there exists any GW (Fig. \ref{Fig:GW_flow}) in this data segment. This work proposes a way to move the detection window on the data stream with some overlapping to increase detection performance. A segment of the data from every window is an above-mentioned sample $\boldsymbol{s}$ and each such sample should be input into the proposed scheme to determine the existence of any GW signal.
	
Furthermore, the experiments of Fig. \ref{fig:Performance:WPCNN_MF} used the matched filtering SNR (equation \ref{Equ:SNR:SNRmf}) which is computed based on a template bank. In this work, this template bank consists of the 25000 GW signals in Training set$^1$ (Table \ref{Tab:DataSets}).

\section{
Principles of the detection scheme}\label{sec:RecScheme}
This section introduces the fundamental principles of the proposed scheme, section \ref{sec:Experiments:Learning} and Table \ref{Table:architecture} present its configurations.

\begin{figure}
  \centering
  \includegraphics[height=3.6cm]{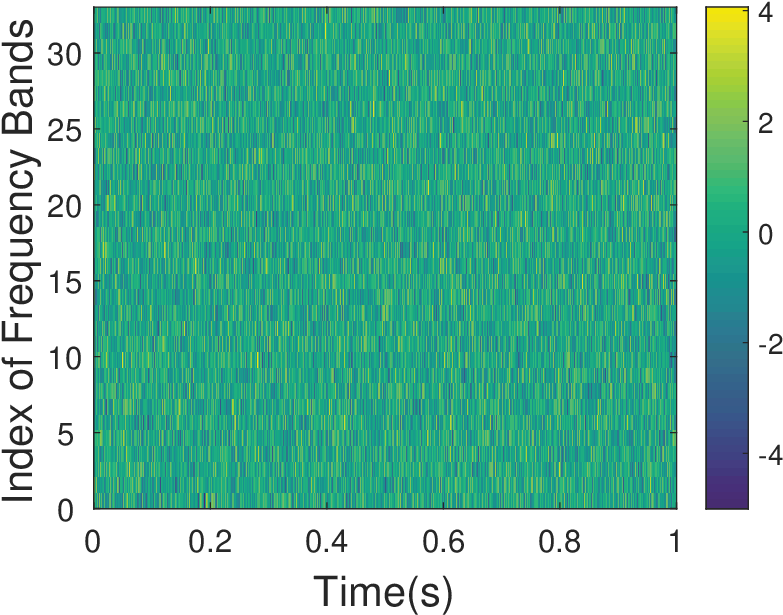}
  \includegraphics[height=3.6cm]{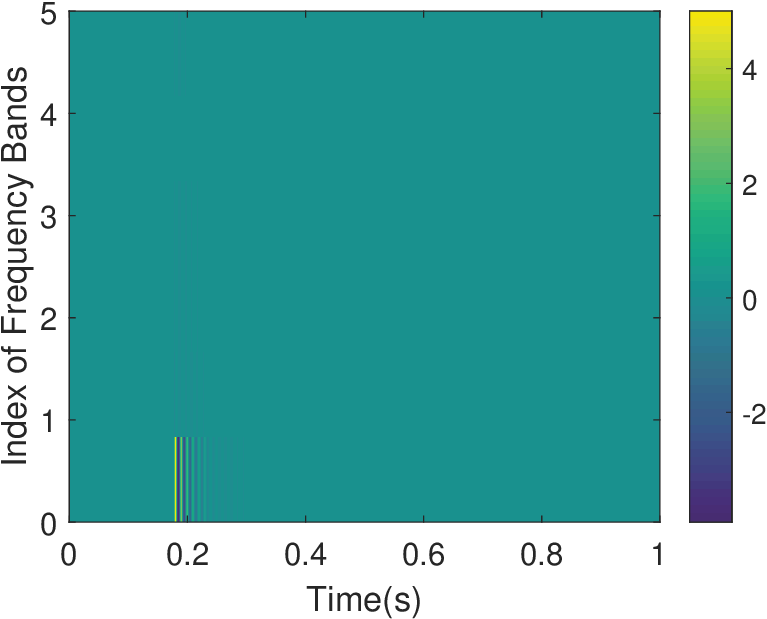}
  \caption{Differences of the noise and a GW signal in a time-frequency space. The two subfigures are the Time-frequency analysis of the noise and GW in Fig. \ref{Fig:GW_flow}, respectively. In this figure, a color indicates the intensity of the corresponding data component. The horizontal axis is time, and the vertical axis is the index of frequency bands in its WP decomposition using Daubechies basis \citep{Book:Mallat:2009}. The frequency in this figure and that of the GW signal are different although there are some positive correlation between them. The GW signal is too weak to be observed in its original time-frequency space. To increase the visibility of the GW signal, therefore, we reduce the range of vertical axis from [0, 32] to [0, 5] in the second subfigure. } \label{fig:GW-WP}
\end{figure}

\subsection{
A Framework of the proposed scheme}\label{sec:RecScheme:Frame}
Theoretically, we could directly use the model in equation (\ref{Equ:estimation_model}) to determine whether there exists a GW signal in $\boldsymbol{s}$. However, it is shown that there is some evident differences  between a GW signal and noises (Fig. \ref{Fig:GW_flow}) in frequency space (Fig. \ref{fig:GW-WP}). Therefore, we can project the observed data $\boldsymbol{s}$ to be analyzed into a frequency space,
	\begin{equation}\label{Equ:g}
	\boldsymbol{z} = g(\boldsymbol{s}),
	\end{equation}
	before determining whether there exist  any GW component in $\boldsymbol{s}$ as follows
	\begin{equation}\label{Equ:estimation_model_2s}
	y=h(\boldsymbol{z})=h(g(\boldsymbol{s})).
	\end{equation}
The $\boldsymbol{z}$ is a representation of $\boldsymbol{s}$ in a frequency space. In this work, the function $g$ is a WP decomposition, and the function $h$ is estimated using a CNN network from a training set.
	
\begin{figure}
        \centering
        \includegraphics[height=1.0cm]{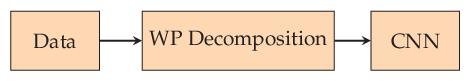}
\caption{A flowchart of the proposed scheme. In this flowchart, the `Data' are the data segments in Fig. \ref{Fig:GW_flow}. WP: wavelet packet, CNN: convolutional neural network.}
\label{fig:flowchart}
\end{figure}

Actually, the scheme based on equations (\ref{Equ:g}) and (\ref{Equ:estimation_model_2s}) is a special implementation of model (\ref{Equ:estimation_model}). This scheme splits model (\ref{Equ:estimation_model}) into two sequential procedures: preprocess the observed data $\boldsymbol{s}$ by $\boldsymbol{z} = g(\boldsymbol{s})$, and
 determine the existence of any GW signal in $\boldsymbol{s}$ by $y=h(\boldsymbol{z})$. This splitting scheme can simplify a complex work to some degree. In machine learning and data mining communities, the procedure $\boldsymbol{z} = g(\boldsymbol{s})$ is called feature extraction. The feature extraction is to simplify the establishing of model (\ref{Equ:estimation_model_2s}) by finding a more approximate representation of the data to be analyzed, $\boldsymbol{s}$ here, and removing irrelevant or weakly-related data components. A flow chart of the proposed scheme is presented in Fig. \ref{fig:flowchart}. We will elaborate  it in the following parts of this section.

In this proposed scheme (Fig. \ref{fig:flowchart}), the WP decomposition is a time-frequency analysis method. Actually, there are a series of investigations on the applications of time-frequency for GW data analysis. For example, \citet{CQG:Chatterji:2004} studied the methods for detecting gravitational-wave bursts based on wavelet transform and Q-transform; \citet{Journal:Klimenko:2008CQG29,Journal:Klimenko:2016PRD19} and \citet{Journal:Abbott:2017:PRD} investigated a coherent Waveburst (cWB) algorithm for detection and reconstruction of gravitational wave bursts; \citet{Journal:Abbott:2017:PRD} and \citet{Journal:Lynch:2017PRL18} researched a framework, omicron-LIB (oLIB), for detecting short-duration gravitational-wave bursts; \citet{Journal:Cornish:2015CQG30}, \citet{Journal:Littenberg:2015PRD31} and \citet{Journal:Abbott:2017:PRD} studied a BayesWave (BW) approach for searching `un-modeled' transient signals; \citet{Journal:Chatterji:2006PRD} and \citet{Journal:Sutton:2010NJP} presented a fully automated search software package, X-Pipeline, for detecting bursts associated with gamma-ray bursts (GRBs) and other astrophysical triggers by coherently analyzing the data from interferometer networks. However, the existing CNN-based investigations for GW signal detection are conducted by inputting a time series into a CNN network without studies on the effects from this kind frequency procedures. Therefore, the necessity of time-frequency analysis and the affects from time-frequency analysis methods are still two open problems to be studied in CNN-based GW detection scheme, and this work studied the WP decomposition for the CNN-based GW signal detection scheme.

Apart from the WP transform, Fourier transform \citep{Howell:2016}, wavelet transform \citep{Journal:Mallat:1989,Book:Mallat:2009,Book:Daubechies:1992} and principal component analysis (PCA) are three typical frequency analysis methods. For example,  \citet{IJMPC:Rampone:2013} investigated the glitch-burst discrimination and glitch classification problems based on PCA and forward neural network (FNN), \citet{CQG:Vinciguerra:2017} studied a scheme to  classify waves emitted in CBC using PCA and an average of FNNs with three hidden layers. As an earlier stage investigation, however, this work focuses on the application of WP decomposition in the CNN-based GW signal detection, and postpones the systematic studies on effects of various frequency analysis methods to future works. Our experiments show that WP is applicable in the GW detection problem and there isn't significant difference on GW detection between typical wavelet bases: Biorthogonal basis (bior), Coiflets (coif), Daubechies basis (db), Haar (haar), ReverseBior (rbio), and Symlets (sym) \citep{Book:Mallat:2009,Journal:Mallat:1989,Book:Daubechies:1992}.

Therefore, we use a WP transform with a db1 base in equation (\ref{Equ:g}). There are multiple variants for basis function db in the implementation of Matlab wavelet toolbox \citep{Matlab:2017}. The number behind db is the index of a specific variant. A brief introduction to the principle and implementation of WP transform can be found in \citet{Li:2015}. That introduction need very little mathematical knowledge to read. For WP decomposition has both time analysis and frequency analysis capabilities, its decomposition result of an observed data (Fig. \ref{Fig:GW_flow}) is an image in a two-dimensional space (Fig. \ref{fig:GW-WP}).

	\subsection{
Convolutional neural networks}\label{sec:RecScheme:CNN}
One typical characteristic of the GW detection problem is that the occurrence time of GW signal is uncertain in the detection window (Fig. \ref{Fig:GW_flow}). After a WP decomposition, a sample to be analyzed is represented using a two-dimensional image with some translation uncertainties (Fig. \ref{fig:GW-WP}). For this kind problem, a typical scheme is the CNN \citep{LECUN:1990,LECUN:1998}. The CNN has been widely investigated in image understanding and computer vision \citep{LECUN:2015}.

\begin{figure}
  \centering
  \includegraphics[height=2.5cm]{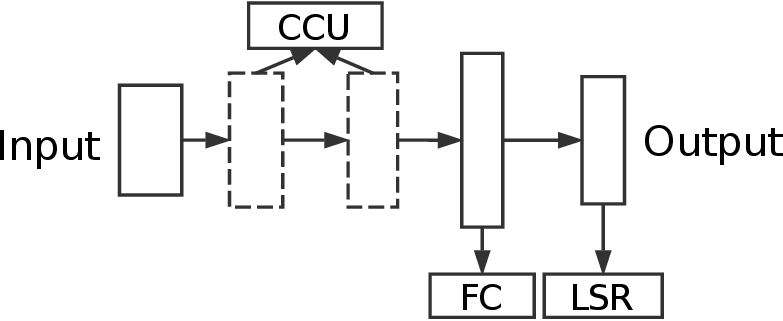}\\
     \vspace{0.3cm}
  \includegraphics[height=2.5cm]{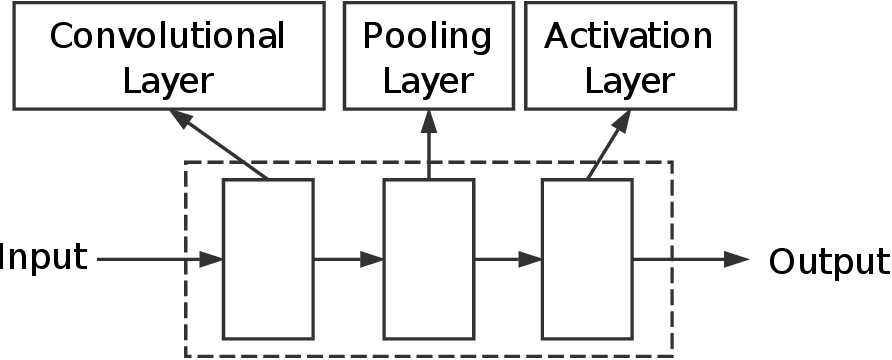}
  \caption{The structure of a convolutional neural network (CNN). The first subfigure: The overall structure of a CNN; The second subfigure: The structure of a composite computing unit (CCU). In this work, we used two CCUs, one fully connected layer (FC) and one logistic regression layer (LSR). Therefore, this work uses two convolutional layers (The second subfigure). Every convolutional layer consists of a series of convolutional kernels (The first subfigure of Fig. \ref{Fig:con_maxpooling}). The configurations of this work's CNN are described in Table \ref{Table:architecture}. }
  \label{Fig:CNN}
\end{figure}

The structure of a CNN used in this wok is presented in Fig. \ref{Fig:CNN}. This CNN network consists of an input layer, several composite computing units (CCUs), a fully connected layer and a logistic regression layer (the first subfigure of Fig. \ref{Fig:CNN}). The input layer receives the WP decomposition (Fig. \ref{fig:GW-WP}) of a data segment to be processed (Fig. \ref{Fig:GW_flow}). A CCU consists of a convolutional layer, a pooling layer, and an activation layer (the second subfigure of Fig. \ref{Fig:CNN}). This work used a CNN with two CCUs.

\begin{figure}
  \centering
  \includegraphics[height=3.4cm]{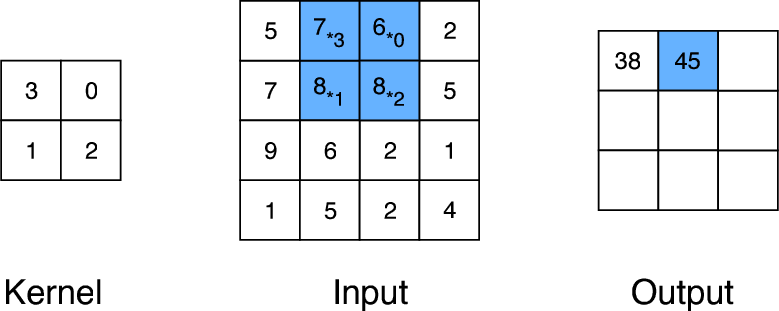}\\
     \vspace{0.3cm}
  \includegraphics[height=3.4cm]{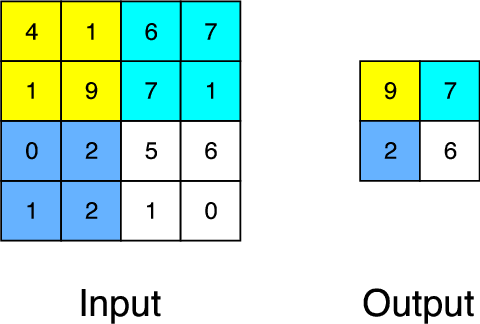}
  \caption{Two sketch maps of a convolutional operation and a max pooling. The first subfigure: The output 38 is computed as $3 \times 5+0 \times 7 + 1 \times 7 + 2 \times 8 = 38$ by aligning the kernel with input at 5, where 3, 0, 1 and 2 come from the kerenl, and 5, 7, 7 and 8 come from the input; the output 45 is computed as $3 \times 7+0 \times 6 + 1 \times 8 + 2 \times 8 = 45$ by aligning the kernel with input at 7. The second subfigure: Max pooling is to obtain an output by applying a max filter to some subregions of the initial representation: output 9 is computed by maximizing  4, 1, 1, 9 (yellow part), 7 computed by maximizing 6, 7, 7 and 1, and so on. The input in the first subfigure is a time-frequency image (Fig. \ref{fig:GW-WP}). This work used two convolutional layers (The second subfigure of Fig. \ref{Fig:CNN}) with 15 convolution kernels and 20 convolution kernels respectively. }
  \label{Fig:con_maxpooling}
\end{figure}

 There are two key concepts, convolution kernel (CK) and pooling for CNN, which represent the computations in the convolutional layer and the pooling layer respectively. A CK  characterizes a discriminant pattern in a problem to be investigated(Fig. \ref{Fig:con_maxpooling}). In case of the data inputted to a convolutional layer being a matrix or a tensor, the CK is defined using a matrix and a tensor respectively, where a tensor is a stack of some matrices. In a convolutional layer, we move a CK on the data to be analyzed, do convolution computation between the CK and the data segment under the CK. By the convolution operation, the presence of some discriminant characteristics of GW signals can be evaluated. The results of a convolutional operation are referred to as convolution responses. The input into a convolutional layer is the output from its previous layer of the neural network. The previous layer can be the input layer of the CNN or an activation layer (Fig. \ref{Fig:CNN}).

In a CNN, multiple CKs can be used for each convolutional layer, the size and number of the CKs on different layers are independent from each other in case of more than one convolutional layers existing. This work uses two convolutional layers, and there are 15 CKs and 20 CKs on these two layers respectively. Furthermore, selections of the CK(s) have some fundamental influences on detection performance, and should be made based on the characteristics of the problem to be investigated. For the knowledge of GW is embodied in training data in this kind statistical schemes, configurations of the CK are learned from training data. The learning method is a back propagation algorithm \citep{Rumelhart:1986}.
	
In the outputs of a convolutional layer, there is usually much redundancy. The existence of redundancy can result in evident degradation of detection performance and computational efficiency on test data. To overcome this kind problem, an operation, pooling, is adopted. The pooling reduces the redundancies from data by merging the convolution responses in every window (the second subfigure of Fig. \ref{Fig:con_maxpooling}), which is referred to as a pooling window. The merging is implemented by computing the maximum of the convolution responses in this pooling window in this work. This pooling window is a 4$\times$4 matrix in this work. More about pooling can be found in \cite{Goodfellow:2016}.

\section{
Experimental evaluations}\label{sec:Experiments}
\subsection{
Learning of the proposed scheme and its specific configuration}\label{sec:Experiments:Learning}
The structure of the GW detection system is introduced in Fig. \ref{fig:flowchart} and Fig. \ref{Fig:CNN}. In this system, the following parameters need to be determined: number of CCUs, number of CKs in every convolutional layer, size and configurations of every CK, size of pooling window and the parameters in the logistic regression layer.

Theoretically, the above-mentioned parameters can be selected based on some optimization theories using training data. However, it is a hybrid, complex optimization problem consisting of both discrete parameters and continuous parameters. For example, the number of CCUs, the number of CKs in every convolutional layer, the size of every CK and the size of a pooling window are discrete parameters, should be positive integers; Configurations of every CK and the parameter of logistic regression are from real space and continuous parameters. Therefore, it is a complex and hybrid optimization problem to learn the proposed scheme.
	
To make its computational complexity be acceptable, the discrete parameters are chosen based on experiences. We used two CCUs, 15 and 20 convolution kernels respectively for the two convolutional layers from the input end to the output end of the CNN (the first subfigure of Fig. \ref{Fig:CNN}). The continuous parameters are estimated using the back propagation algorithm \citep{Rumelhart:1986} from training data (section \ref{sec:data})
	.
	
\begin{figure}
        \centering
        \includegraphics[height=1.86cm]{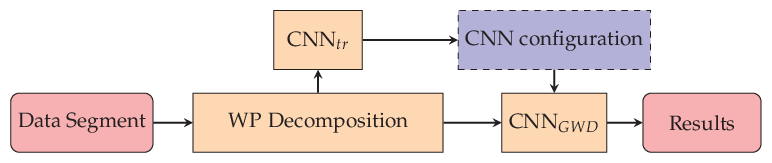}
		\caption{A diagram to show the modules for a software pipeline implementing the proposed scheme and data flow in it. The CNN$_{tr}$ is to learn the configurations of CKs, the connections in step 10 of Table \ref{Table:architecture}, and the parameters  in logistic regression. CNN$_{GWD}$ detects the existence of any GW in a data segment.
}
		\label{fig:software}
	\end{figure}

Some detailed configurations of the proposed scheme are presented in Table \ref{Table:architecture}. The used training sets are described in Table \ref{Tab:DataSets}. Fig. \ref{fig:software} shows the architecture of a proposed implementation for the scheme in this work.

In summary, the input of the proposed scheme (Fig. \ref{fig:software}) is a segment of a time series (Fig. \ref{Fig:GW_flow}), the output indicates whether there exists any GW signal in a detection window that the segment is extracted from. Therefore, if an output implies an occurrence of a GW signal, the time interval of the data segment window is an estimation of the GW occurrence time, and it's not necessary to combine the CNN results from multiple segments together. Otherwise, if we combine the outputs computed from several segments from different windows by maximizing them, then we can only know whether there exist some GW signals in these windows, but lose the number of times GW event occuring and the occurrence time estimation(s).

\begin{table*}
\centering
\caption{The architecture of the proposed scheme. For the experiments in section \ref{Sec:Experiments:Evaluation_Measures}, the inputs come from the training set$^1$ in Table \ref{Tab:DataSets} in learning stage, and test set$^1_1$ (Table \ref{Tab:DataSets}) in the performance evaluation stage. WP: wavelet packet, CK: convolutional kernel, ReLU: Rectified Linear Unit, DSt: data structure. DCI: in the layer with index 3 (Layer 3), the result is a $32\times 126$ matrix from computations based on each CK, the output of Layer 3 is a tensor by stacking the results of 15 CKs; In the output of this layer, there is a third dimension, induced by the index of CKs, which is referred to DCI (Dimension of CK index). The size of the input is 4096.}
\footnotesize
\begin{tabular}{|c|c|c|p{10cm}|}
  \hline
  Index  & Procedures                  &DSt (data size)  & Description                                                                             \\  \hline
  1      & input                       &vector (4096)                    & extract a segment as a sample from a data stream                                        \\  \hline
  2      & WP decomposition            &matrix (32$\times$128)           & conduct a 5-level WP decomposition using a Daubechies basis                             \\  \hline
  3      & convolutional layer         &tensor (32$\times$126$\times$15) & 15 CKs, size of every CK is $1\times3$ (frequency axis$\times$time axis)                                                 \\  \hline
  4      & pooling layer               &tensor (8$\times$32$\times$15)   & size of the pooling window is $4\times 4$                                               \\  \hline
  5      & activation layer            &tensor (8$\times$32$\times$15)   & using the ReLU \citep{Goodfellow:2016} as an activation function\\  \hline
  6      & convolutional layer         &tensor (8$\times$31$\times$20)   &  20 CKs, size of every CK is $1\times2 \times 15$, (frequency axis$\times$time axis $\times$ DCI)                                      \\  \hline
  7      & pooling layer               &tensor (2$\times$8$\times$20)    &  size of the pooling window is $4\times 4$                                              \\  \hline
  8      & activation layer            &tensor (2$\times$8$\times$20)    &  using the ReLU  as an activation function                                              \\  \hline
  9      & flatten layer               &vector (320)                     &  reshape data from a tensor into a vector by stacking its elements                   \\  \hline
  10     & full connected layer        &vector (10)                      &  consists of 10 hidden units with the ReLU as activation function                       \\  \hline
  11     & activation layer            &vector (10)                      &  using the ReLU  as an activation function                                              \\  \hline
  12     & logistic regression layer   &vector (2)                      &  estimate the probability of a sample belonging to each class                               \\  \hline
  \hline
\end{tabular}\label{Table:architecture}
\normalsize
\end{table*}

\subsection{
	Evaluation Measures}\label{Sec:Experiments:Evaluation_Measures}

This work uses sensitivity and a receiver operating characteristic (ROC) curve to evaluate the proposed scheme. The ROC curve is defined by sensitivity and False Positive Rate (FPR). In some literatures, the sensitivity is also referred to as recall, true positive rate (TPR) or true-alarm probability. To give their definitions, we introduce several notations.

Suppose $S$ is a set of observed data,
$$S_1=\{(\mathbf{s}, y): (\mathbf{s}, y) \in S \text{ and there exists GW in } \mathbf{s} \},$$
$$S_2=\{(\mathbf{s}, y): (\mathbf{s}, y) \in S \texttt{ and } (\mathbf{s}, y) \not\in S_1 \},$$
 and $\hat{S}_1$ and $\hat{S}_2$ are respectively the estimations of $S_1$ and $S_2$ based on the proposed GW detection system. That is to say, $S_2$ consists of only noise cases, and each sample in $S_1$ is superimposed by noise and GW. Therefore, the GW detection problem can be dealt with by determining whether an observed data segment belongs to $S_1$ or $S_2$ (equation \ref{Equ:estimation_model}).

Using the above-mentioned notations, the sensitivity and FPR on $S$ are defined as following:
	\begin{equation}\label{Equ:recall}
	\texttt{Sensitivity}(S) = {|S_1\cap \hat{S}_1|}/{|S_1|},
	\end{equation}
	\begin{equation}\label{Equ:FPR}
	\texttt{FPR}(S)  =  \frac{|\hat{S}_1 \cap S_2 |}{|S_2|},
	\end{equation}
where, $|\cdot|$ is to count the number of samples in a data set. Therefore, FPR is an indicator of false alert (false-alarm probability in GW community language).

\subsection{
	Eperimental Evaluations}\label{Sec:Experiments:Experimental_Evaluations}	

In literature, two typical methods for GW detection are matched filtering \citep{Journal:Allen:2012,Journal:Cannon:2012,Journal:Babak:PRD2013,Journal:Usman:2016,Journal:Abbott:2016PRD122004} and CNN  \citep{PRB:George:2018,Journal:Gabbard:2018PRL,ArXiv:Sara:2017,Journal:Zevin:2017}. The matched filtering does GW detection by comparing an observed data with each template in a bank. In comparing evaluation, the matched filtering uses each data element both of noise and interesting GW signal. On the other hand, the CNN is a machine learning method which can establish a mapping from some experiences (training data). This mapping of the CNN with a logistic regression output layer can give a result indicating whether there exists any GW signal in the input data. And the CNN-based scheme in literatures shows comparable detection performance with the benchmark scheme matched filting \citep{PRB:George:2018,Journal:Gabbard:2018PRL,ArXiv:Sara:2017,Journal:Zevin:2017}.

\begin{figure}
		\centering
			\includegraphics[height=4.8cm]{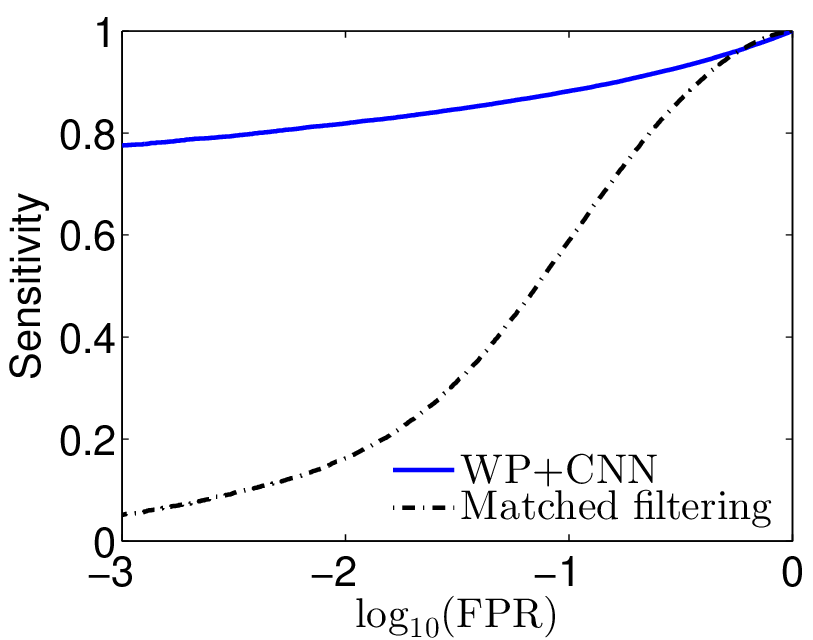}
		\hspace{0in}
		\caption{Performance comparison between the proposed scheme and the matched filtering. The evaluation results in this figure is computed on Test set$^1_1$ (Table \ref{Tab:DataSets}). }
		\label{fig:Performance:WPCNN_MF}
	\end{figure}

In theory, However, the performance of a GW detection system depends on a series of factors, for example, the configuration of detection window, data resolution, time-frequency analysis, and configurations of the CNN. Therefore, this work did some improvements upon these factors based on the pioneering works  \cite{PRB:George:2018} and \cite{Journal:Gabbard:2018PRL}. This paper proposed a detection window with a duration 0.03 seconds and an overlap 0.004 seconds between two successive detection windows. A sample is represented using a series of amplitude values sampled evenly on a time axis in this work. Data resolution (DR) is measured using Hertz (Hz), which refers to the number of pixels/amplitude values per second (pixels are evenly sampled on time axis in this work). This work uses the DR = $1.37 \times 10^5$Hz and wavelet decomposition for time-frequency analysis. The experimental results in Fig. \ref{fig:Performance:WPCNN_MF} show the effectiveness of these improvements.

\begin{figure}
		\centering
            \includegraphics[height=4.8cm]{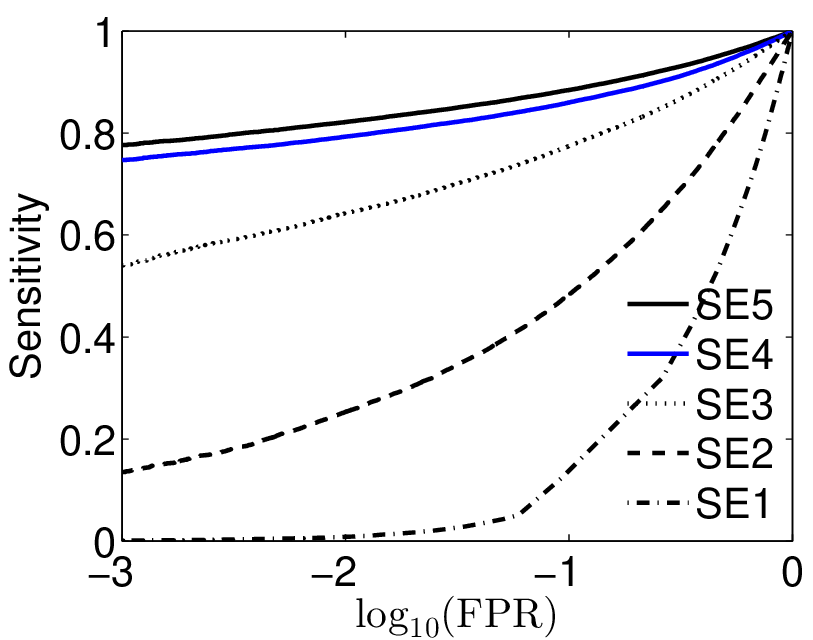}
		\hspace{0in}
		\caption{ROC curve: a comprehensive evaluation on the performance of the proposed improvement schemes. This evaluation was conducted by gradually changing the detection scheme configuration from that of \citet{Journal:Gabbard:2018PRL} to this work's proposal (Table \ref{Table:architecture} and section \ref{sec:Experiments:Learning}), the detailed configurations of this experiment are presented in Table \ref{Table:architecture:comprehensive_evaluation}. In this experiment, the SE1 was learned from Training set$^2$ (Table \ref{Tab:DataSets}), evaluated on Test set$^2_1$ (Table \ref{Tab:DataSets}), and SE$i$ ($i =2, 3, 4, 5$) were learned and tested on Training set$^1$ and Test set$^1_1$ respectively (Table \ref{Tab:DataSets}).
		}
		\label{fig:Performance:comprehensive_evaluation}
	\end{figure}

To observe the effectiveness of each improvement individually, \textbf{this work conducted a comprehensive evaluation by gradually changing the detection scheme from that of \citet{Journal:Gabbard:2018PRL} to this work's framework} (Table \ref{Table:architecture} and section \ref{sec:Experiments:Learning}) (Thank H. Gabbard, M. Williams, F. Hayes, \& C. Messenger very much for their enthusiastic discussions and sharing of their experimental code). The configuration and results of this evaluation experiment are presented in Table \ref{Table:architecture:comprehensive_evaluation}, Fig. \ref{fig:Performance:comprehensive_evaluation} and Table \ref{Tab:Performance:comprehensive_evaluation} respectively.

In this evaluation, experiment SE1 utilize the configurations of \citet{Journal:Gabbard:2018PRL}, for example, CNN structure, data resolution, detection window. The experiments SE2, SE3, SE4, SE5 test the performance of CNN structure \& detection window width optimization, data resolution, wavelet packet decomposition and detection window overlap scheme (Table \ref{Table:architecture:comprehensive_evaluation}). The results in Table \ref{Tab:Performance:comprehensive_evaluation} and Fig. \ref{fig:Performance:comprehensive_evaluation} show that the proposed optimization scheme have evident improvement on detection performance. This work and \citet{Journal:Gabbard:2018PRL} used different theoretical model for simulating gravitation wave signal. Therefore, there exist some differences between the detection results of SE 1 and \citet{Journal:Gabbard:2018PRL}. After these optimizing strategies on time-frequency analysis, detection window, data resolution and pooling scheme, the CNN-based method shows some evident potentials to detect the GW much more sensitively than the matched filtering (Fig. \ref{fig:Performance:WPCNN_MF}).

Comparing with the matched filtering method, one remarkable characteristic of the proposed scheme is its efficiency \citep{PRB:George:2018,Journal:Gabbard:2018PRL}. Although the learning of the configuration for a CNN is the most time-consuming stage in the proposed scheme, it can be done beforehand. In test stage, we only need to compute the output of a CNN using the learned parameters. However, matched filtering method will evaluate the similarity of every sample to be checked with each template in a bank. For convenience, this characteristic of matched filtering is referred to as every-template-evaluation-for-each-test-sample (ETEETS). To have an accurate detection performance, the bank should span a large astrophysical parameter space and consist of many samples/templates. The huge volume of the bank and the ETEETS characteristic result that the matched filtering method is computationally inefficient. The experiments of this work show that the proposed CNN-based scheme takes 18.765 milliseconds to process a sample on a computer using twelve Inter(R) Xeon(R) CPUs (E7-4830) with frequency 2.31GHz, while the matched filtering 1899.350 milliseconds to process a sample. One additional note, in the experiment SE 5 (Fig. \ref{fig:Performance:comprehensive_evaluation}), the training time of the proposed CNN model is approximately 430 minutes, and the Wavelet Packet decomposition time is approximately 25 minutes for the training set$^1$.

\begin{table}
\centering
\caption{The sensitivities in case of FPR= 0.001, 0.01, 0.1 and $\rho_{opt}=2$. Considering the lowest SNR $\rho_{opt}$ of the test data in two closely related works \citep{PRD:George:2018,PRB:George:2018} is 2, the Table 3 presents the evaluation results on the data with $\rho_{opt}=2$ from Test set$^2_1$ and Test set$^1_1$. That is to say, the test data in experiments of this table is a subset of those of Fig. \ref{fig:Performance:comprehensive_evaluation}.
      Exp. No.: Experimental identification. Some detailed configurations of this experiments are presented in Table \ref{Table:architecture:comprehensive_evaluation}.}
      \footnotesize
	\begin{tabular}{ccccc}
		\hline
		\hline
		Exp. No. & FPR=0.001 & FPR=0.01 & FPR=0.1\\
		\hline
		SE1 & 0.001 & 0.005 & 0.300 \\
		SE2 & 0.067 & 0.365 & 0.555 \\
		SE3 & 0.660 & 0.748 & 0.835\\
		SE4 & 0.853 &  0.883 & 0.918\\
		SE5 & 0.947 & 0.967 & 0.984\\
		\hline
	\end{tabular}\label{Tab:Performance:comprehensive_evaluation}
\normalsize
\end{table}

\begin{table*}
\centering
\caption{Configurations of the comprehensive evaluation experiments on the proposed optimization schemes. WPD: wavelet packet decomposition, DS: data structure,
CK: convolutional kernel, ReLU: Rectified Linear Unit \citep{Goodfellow:2016}, ELU: exponential linear unit, configuration of a convolutional layer $(i_1, i_2, \cdots, i_k)$: $i_1$ is convolutional kernels (CKs) and $i_2 \times \cdots \times i_k$ ($k= 3$ or $4$) is size of every CK , configuration of a
pooling layer $(j_1, j_2)$: size of the pooling window is $j_1 \times j_2$, configuration of a fully connected layer $i$: $i$ is the number of hidden units with the ReLU as activation function. In experiment SE1, there isn't any overlap between two successive pooling window as proposed in \citet{Journal:Gabbard:2018PRL}, but there exist some overlaps in experiments SE$i$ ($i\geq 2$).
}
\footnotesize
\begin{tabular}{|c|c|c|c|c|c|c|}
  \hline
  Procedures                 &SE1             &SE2             &SE3              &SE4               &SE5              \\ \hline
  \multicolumn{6}{|c|}{Optimization schemes}                                                                                         \\  \hline
  data resolution (Hz)       &8192            &8192             &1.37$\times 10^5$ &1.37$\times 10^5$ &1.37$\times 10^5$\\  \hline
  window width               &1s              &0.03s            &0.03s             &0.03s             &0.03s            \\  \hline
    WP decomposition           &No              &No               &No                &Yes               & Yes             \\  \hline
  window overlap             &No              &No               &No                &No                & 0.004s          \\  \hline
    \multicolumn{6}{|c|}{Configurations of CNN(based on the order from input layer to output layer)}                                 \\ \hline
   convolutional layer       &(8, 1, 64)      &(15, 1, 3)       &(15, 1, 3)        &(15, 1, 3)        & (15, 1, 3)      \\  \hline
   pooling layer             &No              &\textbf{(1, 16)} &\textbf{(1, 16)}  & (4, 4)           & (4, 4)          \\  \hline
   activation  layer         &ELU             &ReLU             &ReLU              &ReLU              &  ReLU           \\  \hline
   convolutional layer       &(8, 1, 32, 8)   &(20, 1, 2, 15)   &(20, 1, 2, 15)    &(20, 1, 2, 15)    &  (20, 1, 2, 15) \\   \hline
   pooling  layer            &(1, 8)          &\textbf{(1, 16)} &\textbf{(1, 16)}  &(4, 4)            &  (4, 4)         \\  \hline
   activation layer          &ELU             &ReLU             &ReLU              &ReLU              &  ReLU           \\  \hline
   convolutional layer       &(16, 1, 32, 8)  &No               &No                &No                &No               \\ \hline
   activation layer          &ELU             &No               &No                &No                &No               \\ \hline
   convolutional layer       &(16, 1, 16, 16) &No               &No                &No                &No               \\ \hline
   pooling  layer            &(1, 6)          &No               &No                &No                &No               \\ \hline
   activation layer          &ELU             &No               &No                &No                &No               \\ \hline
   convolutional layer       &(32, 1, 16, 16) &No               &No                &No                &No               \\ \hline
   activation layer          &ELU             &No               &No                &No                &No               \\ \hline
   convolutional layer       &(32, 1, 16, 32) &No               &No                &No                &No               \\ \hline
   pooling  layer            &(1, 4)          &No               &No                &No                &No               \\ \hline
   activation layer          &ELU             &No               &No                &No                &No               \\ \hline
   flatten  layer            &Yes             &Yes              &Yes               &Yes               &Yes              \\  \hline
   full connected layer      &64              &10               &10                &10                &10               \\  \hline
   activation                &ELU             &ReLU             &ReLU              &ReLU              &ReLU             \\  \hline
   full connected layer      &64              &No               &No                &No                &No               \\ \hline
   activation  layer         &ELU             &ReLU             &ReLU              &ReLU              &ReLU             \\  \hline
   logistic regression layer &Yes             &Yes              &Yes               &Yes               &Yes              \\  \hline
  \hline
\end{tabular}\label{Table:architecture:comprehensive_evaluation}
\normalsize
\end{table*}

\section{
Conclusions and discussions}\label{sec:Conclusions}

This work investigated the optimization techniques for the CNN-based GW detection scheme and proposed a GW detection scheme based on CNN and wavelet packet decomposition method. The motivation is to fill in the lack of the research on the dependencies of CNN-based GW detection scheme on such vital factors as frequency analysis, configurations of detection window, and data resolution, etc \citep{NIPS:Gebhard:2017,PRD:George:2018,PRB:George:2018,Journal:Gabbard:2018PRL,Journal:Gebhard:PRD2019}.  Experimental results show that the proposed optimization of this work improved the GW detection sensitivity evidently. Therefore, the accumulation effects from such factors could't be ignored in GW detection.

This proposed scheme consists of three essential procedures: firstly, extract a segment from a data stream, secondly, decompose this segment using WP; thirdly, detect the existence of any GW signal using a CNN. It is shown that the WP spotlights the characteristics of GW in a time-frequency space, and can improves the detection performance. In reality, the observed data comes as a data stream. Therefore, this work proposes an overlapping window scheme, by which the GW detection problem is reformulated as a classification paradigm in machine learning.

Actually, this is a very flexible scheme as a reference for related scientists in this field. For example, if preparing some training samples for one or more classes of glitches and replacing the logistic regression with a softmax regression, the proposed scheme can be explored in glitch detection; if replacing the logistic regression with
a linear regression, we can study the application of the proposed scheme in GW parameter estimation in theory, for example, parameters $f_0$ and $\tau$ in equation (\ref{Equ:GW_Model}).

However, the principle of the CNN indicates that there exists more error in estimating parameter $t_0$ in GW model of equation (\ref{Equ:GW_Model}), the occurrence time of a GW in a data steam, than any other parameter. This error comes from the time translation invariance effect of the pooling operation.  The time translation invariance effect means that a small change of gravitational wave occurrence time in the detection window will not affect the calculation results of CNN. Therefore, the significance of this error depends on the width of the detection window. To alleviate the problem, we can firstly estimate parameter $t_0$ using a CNN-based scheme with an optimized window width and obtain an initial estimation, then reestimate it using an CNN method with a shorter detection window around the initial estimation.

When this paper was under review, we became aware of the paper \citep{arXiv:Wang:2019} which conducted an interesting investigation by introducing a novel observation data decomposition method based on theoretical waveforms. In principle, the theoretical waveforms in \citet{arXiv:Wang:2019} play a similar role with the wavelet basis in the proposed scheme of this work. The theoretical waveforms have attractive physical background, the wavelet packet decomposition has concrete theoretical foundations and widely applications. However, it is an interesting topic to study the existence of their different characteristics on the detection completeness of GW events, false alarm rate.

\acknowledgements{
Authors thank H. Gabbard, M. Williams, F. Hayes, \& C. Messenger very much for their enthusiastic discussions and sharing of their experimental code. We are grateful for valuable suggestions and corrections from anonymous reviewers, Eric D. Feigelson, Dr. Jin Li and B.S. Sathyaprakash.
    X. L. and W. Y. are supported by the National Natural Science Foundation of China (grant Nos 11973022, U1811464), the Natural Science
	Foundation of Guangdong Province (2020A1515010710), and China Scholarship Council (201706755006), and the Joint Research Fund in Astronomy (U1531242) under cooperative agreement between the National Natural Science Foundation of China (NSFC) and Chinese Academy of Sciences (CAS).
    Xilong Fan is supported by the National Natural Science Foundation of China (grant Nos 11673008,11922303), and Hubei province Natural Science Fund for the Distinguished Young Scholars.
}


\raggedend

\begin{small}

\end{small}


\end{document}